# Taxonomizing Synthetic Data for Law

*Ignacio Cofone,*[*] *Katherine J. Strandburg,*[**] *and Nicholas Tilmes*[***]

*ABSTRACT: Synthetic data is increasingly important in data usage and AI design, creating novel legal and policy dilemmas. All too often, discussions of synthetic data treat it as entirely distinct from "real," collected data, overlooking the risks posed by different kinds and uses of synthetic data. This piece comments on Michal Gal and Orla Lynskey's work, which persuasively argues that synthetic data will transform information privacy, market competition, and data quality. While the risks posed by synthetic data depend on its connection to collected data, we argue that background knowledge and assumptions about ground truth used to create it are at least as important. We bring that focus to Gal and Lynskey's taxonomy of synthetic data, arguing that it is essential to grasp synthetic data's legal and policy implications. As such, we divide synthetic data into (1) transformed data, which modifies collected data to preserve certain statistical properties for an end use; (2) augmented data, which relies on assumptions to bolster a collected dataset's fidelity to the ground truth; and (3) simulated data, which relies almost entirely on background knowledge and ground-truth assumptions. As policymakers weigh whether to incentivize, mandate, or discourage the use of synthetic data, they should consider the validity of the ground-truth assumptions used in producing that data.*

[*] Professor of Law and Regulation of AI, University of Oxford, Faculty of Law and Institute for Ethics in AI. ignacio.cofone@law.ox.ac.uk.

[**] Pauline Newman Professor of Law, NYU School of Law. katherine.strandburg@nyu.edu.
Professor Strandburg acknowledges the generous support of the Filomen D'Agostino and Max E. Greenberg Research Fund. This material is also based in part upon work supported by the National Science Foundation under Award No. 2131532.

[***] JD, NYU School of Law 2025. net266@nyu.edu.

## Introduction

Synthetic data is becoming a key part of artificial intelligence ("AI") development. In their excellent article *Synthetic Data: Legal Implications of the Data-Generation Revolution*, Michal Gal and Orla Lynskey explore how synthetic data is set to revolutionize data usage, especially in AI.[1] Synthetic data is artificially generated but retains analytical value because it is created using methods intended to represent aspects of ground truth in the real world. Gal and Lynskey's instructive treatment introduces the under-studied issue of synthetic data, including how and why it is created, to a legal audience, and provides an illuminating analysis. They persuasively argue that the shift toward synthetic data requires a re-evaluation of current law because current law is primarily designed to address collected data. Their article focuses on the implications of synthetic data for market dynamics, privacy, and data quality, arguing it will disrupt established competitive advantages and require new regulatory approaches to balance the utility of data applications with social and legal values such as privacy. Gal and Lynskey's contribution is of enormous value to legal scholarship and policymaking.

Legal and policy discussions often draw a sharp line between collected and synthetic data, but the relevance of this distinction to regulatory goals can be over-stated. What matters most for regulatory design is not whether data is "real" or synthetic, but what risks uses of it create for social values such as privacy, accuracy, and equity. The question for regulators is whether and how synthetic data mitigates or exacerbates those risks.

1. *See* Michal S. Gal & Orla Lynskey, *Synthetic Data: Legal Implications of the Data-Generation Revolution*, 109 Iowa L. Rev. 1087, 1154–56 (2024).

In this Comment, while agreeing with Gal and Lynskey that the risks posed by synthetic data depend on the extent to which it is based on collected data, we argue that these risks depend at least as much on the background knowledge and assumptions about ground truth relied upon in creating synthetic data. Focusing on the role of background knowledge and ground-truth assumptions in synthetic data is crucial to distinguishing between types of synthetic data in a way that can successfully address legal and policy considerations and further Gal and Lynskey's normative objectives. This Comment thus aims to bring the importance of background knowledge and ground-truth assumptions to the fore.

Our Comment proceeds as follows. In Part I, we briefly review synthetic data and its applications, then discuss three categories of synthetic data, emphasizing the importance of background knowledge and ground-truth assumptions for evaluating the policy implications of each category. In Parts II, III, and IV we develop the implications of ground-truth assumptions for each of Gal and Lynskey's treatments of legal and policy concerns. We examine privacy law in Part II, data quality in Part III, and competition law in Part IV.

## I. A TAXONOMY OF SYNTHETIC DATA

### A. *WHAT IS SYNTHETIC DATA?*

Synthetic data is artificially generated data that attempts to closely mimic the statistical properties and characteristics of real-world, "ground-truth" data.[2] It serves as a surrogate for or supplement to data collected from the real world, attempting to overcome practical and legal limitations associated with collecting or using "real" data.[3] Often, synthetic data is used to replace personal data, as an alternative to more traditional—and frequently criticized—forms of de-identification.[4] Synthetic data is also increasingly used

2. *See, e.g.*, Zhenchen Wang, Barbara Draghi, Ylenia Rotalinti, Darren Lunn & Puja Myles, *High-Fidelity Synthetic Data Applications for Data Augmentation*, *in* 26 DEEP LEARNING 145, 147 (Manuel Domínguez-Morales, Javier Civit-Masot, Luis Muñoz-Saavedra & Robertas Damaševičius eds., 2024) ("Synthetic data are artificial data that can mimic the statistical properties, patterns, and relationships observed in real-world data."); Yingzhou Lu et al., Machine Learning for Synthetic Data Generation: A Review 5 (Apr. 4 2025) (unpublished manuscript) (on file with the *Iowa Law Review*) ("Synthetic data refers to computer-generated information that mimics the properties of real-world data without disclosing any personally identifiable information.").

3. Gal & Lynskey, *supra* note 1, at 1092–93. *But see generally* Georgi Ganev, *Synthetic Data, Similarity-Based Privacy Metrics, and Regulatory (Non-)Compliance*, 2D WORKSHOP ON GENERATIVE AI & L. 2024 (preprint) (finding that some forms of AI-generated synthetic data fail to several tests for compliance with personal data protection regulations such as the GDPR).

4. *See, e.g.*, Fan Liu et al., *Privacy-Preserving Synthetic Data Generation for Recommendation Systems*, SIGIR '22: PROC. 45TH INT'L ACM SIGIR CONF. ON RSCH. & DEV. INFO. RETRIEVAL 1379, 1388 (2022); Mikel Hernandez, Gorka Epelde, Ane Alberdi, Rodrigo Cilla & Debbie Rankin, *Synthetic Data Generation for Tabular Health Records: A Systematic Review*, 493 NEUROCOMPUTING 28, 28–29 (2022).

to train and test machine learning models in a wide array of domains such as image recognition and autonomous vehicles.[5]

Gal and Lynskey define synthetic data as "artificial data, generally generated by computer simulations or algorithms, which has analytical value."[6] A report commissioned by the Royal Society defines synthetic data somewhat more specifically as "data that has been generated using a purpose-built mathematical model or algorithm, with the aim of solving a (set of) data science task(s)."[7] In this Comment, we discuss synthetic data utilizing the Royal Society definition, meaning that we assume that synthetic datasets are ordinarily generated with specific tasks in mind and that those tasks are ordinarily "data science" tasks. We take this to mean that synthetic data is generally created for tasks that require large datasets, such as training machine learning models.

Synthetic data can be instrumental in training and testing machine learning models as long as it reduces the need to collect large, sensitive, or proprietary sets of "real" data.[8] It can be also used to create profiles of normal behavior, which facilitates the identification of anomalies or cybersecurity threats.[9] In software development, synthetic data can be used to simulate user interactions, helping identify issues and vulnerabilities before real users find them.[10] Most relevant to our discussion here, synthetic data can be used to enable researchers to perform analyses without using identifiable (e.g., patient or financial) information;[11] to augment collected data for purposes such as reducing bias and overcoming statistical imbalances in collected data;[12] or as a more inexpensive and accessible means to accomplish various data-driven tasks.[13] Synthetic data can be generated using different methods, each with its own advantages and drawbacks, including randomization, generative models, data masking, and simulation.[14] While Gal and Lynskey also discuss "curat[ed]" data that has been subject to routine corrections, such as changing minutes to hours,[15] we do not include it in our taxonomy. Data curation, in our view, aims to produce a more accurate or well-formatted set of collected data, rather than to create artificial data to replace or supplement collected data.[16] Thus, while data curation can raise important issues—for example, whether an apparent "outlier" is an error or a rare, but real, event—those issues are not central to the questions about synthetic data addressed here.

---

5. Lu et al., *supra* note 2, at 2–6; Celso M. de Melo et al., *Next-Generation Deep Learning Based on Simulators and Synthetic Data*, 26 TRENDS COGNITIVE SCIS. 174, 174 (2022).

6. Gal & Lynskey, *supra* note 1, at 1090.

7. JAMES JORDON ET AL., SYNTHETIC DATA – WHAT, WHY AND HOW? 5 (2022).

8. Lu et al., *supra* note 2, at 1; de Melo, *supra* note 5, at 174.

9. Garima Agrawal, Amardeep Kaur & Sowmya Myneni, *A Review of Generative Models in Generating Synthetic Attack Data for Cybersecurity*, ELECS., Jan. 2024, at 1, 15–16; Ayesha Siddiqua Dina, A.B. Siddique & D. Manivannan, *Effect of Balancing Data Using Synthetic Data on the Performance of Machine Learning Classifiers for Intrusion Detection in Computer Networks*, 10 IEEE ACCESS 96731, 96733 (2022).

10. Agrawal et al., *supra* note 9, at 19–21.

11. Hernandez et al., *supra* note 4, at 39; Debbie Rankin et al., *Reliability of Supervised Machine Learning Using Synthetic Data in Health Care: Model to Preserve Privacy for Data Sharing*, JMIR MED. INFORMATICS, July 2020, at 485, 486.

12. Shubham Sharma et al., *Data Augmentation for Discrimination Prevention and Bias Disambiguation*, PROC. AAAI/ACM CONF. AI, ETHICS & SOC'Y 358, 358–59 (2020); Nikita Jaipuria et al., *Deflating Dataset Bias Using Synthetic Data Augmentation*, 2020 IEEE/CVF CONF. ON COMPUT. VISION & PATTERN RECOGNITION WORKSHOPS 3344, 3345 (2020).

13. Gal & Lynskey, *supra* note 1, at 1114.

For illustration purposes, we discuss below an infamous example, also discussed by Gal and Lynskey, in which Amazon sought to design an algorithm to rank job applicants. However, "[t]he algorithm was trained on 'resumes submitted to the company' in previous years, which reflected male dominance in the industry. The result was that it judged male applicants as superior, and penalized references in resumes which indicated the applicant was a woman (e.g., women's football captain)."[17] Here, the collected dataset (resumes of previous successful applicants) was incomplete in that it did not include enough examples of resumes from women to allow Amazon to identify likely successful female applicants.[18] More than that, the algorithm apparently learned from the predominance of men's resumes in the collected resumes to identify and reject women who applied. Hypothetically, synthetic data could have been used to improve this situation.

### *B. Synthetic Data Through a Ground-Truth Lens*

Gal and Lynskey helpfully categorize synthetic data according to whether it is produced by methods that (i) transform collected data; (ii) reduce the need for collected data; or (iii) do not (directly) use collected data.[19] Their

---

14. *See, e.g., id.* at 1095–101; Lu et al., *supra* note 2, at 13; JORDON ET AL., *supra* note 7, at 24–27.

15. Gal & Lynskey, *supra* note 1, at 1095 ("[A]s collected data is cleaned and prepared for analysis, some of the values in the dataset are replaced with synthetic values to ensure consistency (a process called data curation). For instance, if most data points relate to minutes, those that relate to hours can be replaced by synthetic data to correct the inconsistency." (footnote omitted)); *see also id.* at 1109 (listing "overcoming business constraints []such as trade secrets or data security[] or legal ones []such as privacy regulation[]" as other uses of synthetic data).

16. *See* Larysa Visengeriyeva & Ziawasch Abedjan, *Anatomy of Metadata for Data Curation*, J. DATA & INFO. QUALITY 9 (Sept. 2020), https://dl-acm-org.proxy.lib.uiowa.edu/doi/pdf/10.1145%2F3371925 [https://perma.cc/P9VJ-6QN8].

17. Gal & Lynskey, *supra* note 1, at 1099–100 (footnote omitted); *see also* Jeffrey Dastin, *Insight - Amazon Scraps Secret AI Recruiting Tool That Showed Bias Against Women*, REUTERS (Oct. 10, 2018, 7:50 PM), https://www.reuters.com/article/idUSKCN1MK0AG (on file with the *Iowa Law Review*) ("In effect, Amazon's system taught itself that male candidates were preferable.").

18. *See* Gal & Lynskey, *supra* note 1, at 1099–100.

19. *Id.* at 1095–101 (identifying "categories, which are distinguished by their need for collected data . . . in the generation process. This parameter enables us to explore the extent to taxonomy highlights distinctions based on the extent to which synthetic data was *derived from collected data.* We argue that this insight must be combined with an emphasis on the ways in which synthetic data *depends on background knowledge or assumptions about the principles and rules that underlie the "ground truth"* that the data purports to represent. (For brevity, we will refer to this kind of background knowledge and assumptions about ground truth as "ground-truth assumptions.").

We argue that a synthetic dataset's reliance on ground-truth assumptions is critical for its uses and limitations. When synthetic data is derived from reliable ground-truth assumptions, it can reflect key properties that would have characterized a large and representative sample of data collected from the real-world, data-generating process, such as higher-order statistical properties, conditional dependencies, and edge cases and rare events that might have occurred. Synthetic data produced by simulating reliable ground-truth assumptions (such as the rules of a game or the laws of gravity) can be reliable even when based on minimal or no collected data. Conversely, synthetic data created using unreliable ground-truth assumptions can be significantly misaligned with real-world behavior, even if it is based directly on collected data. A framework based on reliance on ground-truth assumptions also helps to surface potential harms from synthetic data because the data's risks, for example for privacy and fairness, are directly related to the reliability of the ground-truth assumptions used to create it, as we explain in the following Sections. Building on Gal and Lynskey's work, we propose that regulatory design should be guided by the following synthetic data taxonomy: (i) transformed data; (ii) augmented data; and (iii) simulated data, where the analysis of each category should emphasize the distinctive implications of ground-truth assumptions.

### 1. Transformed data

The first category encompasses synthetic data created by methods that transform collected data based on assumptions about *which of the collected dataset's statistical properties should be preserved* for an anticipated end use. This category corresponds to what Gal and Lynskey call "[g]eneration [b]ased on [t]ransformations of [c]ollected [d]ata."[20] As Gal and Lynskey explain, transformation methods rely on a selection of observations about the collected dataset, such as its "(main) statistical characteristics," to create "a synthetic [dataset] with quite similar characteristics."[21] Because such transformations reduce fidelity to the collected dataset, they are ordinarily which data collection barriers can be overcome, as well as the potential use of private data in the generation process.").

20. *Id.* at 1095.

21. *Id.* at 1096.

deployed to hide details of the collected dataset for purposes such as cybersecurity, privacy, or trade secrecy.[22]

Transformed data includes synthetic data created by various methods, such as randomizing existing data by adding noise to it while preserving its statistical properties.[23] For example, in healthcare, patients' ages or medical test results can be perturbed within a range to create synthetic data that maintains the original data's distribution.[24] Other approaches, such as differential privacy and k-anonymity, seek to preserve more of the data's utility by adding noise in a carefully prescribed manner. Differential privacy is designed to ensure that the presence or absence of an individual's data does not significantly impact the outcome, making it more difficult for attackers to infer specific information about individuals.[25] K-anonymity generalizes or suppresses data so that each record in the dataset is indistinguishable from (at least k-1) other records with respect to some attributes.[26]

The distinctive aspect of transformed synthetic data is that it is intended to replace collected data *while not aiming to improve the collected dataset's fidelity to ground truth.* Indeed, this kind of synthetic data is ordinarily *farther* from ground truth than the collected data it is based on because it inherits any limitations of the initial sampling process (such as unrepresentativeness), while distorting the data further through some form of randomization. This degradation of fidelity trades off with the potential benefits of improving or preserving privacy, cybersecurity, or trade secrecy.

In the Amazon example, suppose the company wanted to use crowdsourcing to help it understand what went wrong with its hiring algorithm and design a less biased algorithm. It might transform gendered aspects of the resume data to preserve prior applicants' privacy before sending the data out to the crowd for evaluation. Netflix famously used a transformation approach like this to de-identify data in its movie ranking database in 2007, which failed spectacularly (although perhaps doing an excellent job of preserving the important correlations between individuals and rankings).[27] In retrospect, Netflix might have done better to sacrifice fidelity to the collected data in exchange for better privacy protection.

---

22. *See generally* Ahmed M. Alaa, Boris van Breugel, Evgeny Saveliev & Mihaela van der Schaar, *How Faithful is Your Synthetic Data? Sample-Level Metrics for Evaluating and Auditing Generative Models*, 162 PROC. 39TH INT'L CONF. ON MACH. LEARNING 290 (2022) (explaining the process for developing evaluation for generative models); Rankin et al., *supra* note 11 (accepting small decreases in accuracy from models trained with synthetic data due to lessened privacy risks); JORDON ET AL., *supra* note 7 (providing an overview of the state of play of synthetic data).

23. Yuzheng Hu et al., *SoK: Privacy-Preserving Data Synthesis*, 2024 IEEE SYMP. ON SEC. & PRIV. 4696, 4698 (2024) (listing synthetic data generation techniques that integrate learning with perturbation).

24. *Id.*; Rankin et al., *supra* note 11, at 486 ("[D]ata perturbation techniques such as data swapping, data masking, cell suppression, and adding noise have been applied to real data to modify and thus protect the data from disclosure prior to releasing it.").

25. *See* CYNTHIA DWORK & AARON ROTH, THE ALGORITHMIC FOUNDATIONS OF DIFFERENTIAL PRIVACY 5–6 (2014) (describing the original purpose of differential privacy).

26. *See generally* Latanya Sweeney, *k-Anonymity: A Model for Protecting Privacy*, 10 INT'L J. ON UNCERTAINTY, FUZZINESS & KNOWLEDGE-BASED SYS. 557 (2002) (being the original work for k-anonymity).

### 2. Augmented data

The second category covers synthetic data generated by methods that rely on *assumptions about the underlying ground truth* to *augment the collected dataset* with the goal of improving its fidelity to ground truth for a particular purpose. Gal and Lynskey describe these methods as those that "[r]educe the [n]eed for [c]ollected [d]ata."[28] Our ground-truth argument emphasizes that whether the need for collected data is reduced depends on the reliability of the ground-truth assumptions underlying the augmentation.[29]

Various methods can be used to create augmented synthetic data. For example, consider Gal and Lynskey's example of an AI image generator that augments a dataset of stop sign pictures by iteratively "determin[ing] that [a generated] image is fake based on mismatches between the fake image and collected data available to the algorithm" and rejecting images that the algorithm recognizes as fake.[30] These creation and assessment processes require assumptions about how well the collected data represents the ground truth, such as the regularity of physical objects near stop signs, and knowledge about natural phenomena such as light and shadow. Generative models, such as Generative Adversarial Networks ("GAN") and variational autoencoders, use machine learning to capture underlying patterns in collected data and generate synthetic samples that closely resemble them.[31] GANs have become popular because of their ability to create realistic images, text, and audio. This process treats a collected dataset as a random sample of a hypothetical ground-truth population to which the model generalizes based on assumptions about the hypothetical ground-truth distribution.[32] Thus, the goal is to create additional data that is consistent with the available collected data—in the spirit of interpolation, rather than extrapolation. Depending on the reliability of the data collected and the ground-truth assumptions employed, using this method to augment a dataset may or may not reduce the need for additional collected data.

---

27. Bruce Schneier, *Why 'Anonymous' Data Sometimes Isn't*, WIRED (Dec. 12, 2007, 9:00 PM), https://www.wired.com/2007/12/why-anonymous-data-sometimes-isnt [https://perma.cc/9JUP-ZK46].

28. Gal & Lynskey, *supra* note 1, at 1098.

29. We return to this point later in this Comment because it has important implications for the discussion of legal reforms.

30. Gal & Lynskey, *supra* note 1, at 1099.

31. Lu et al., *supra* note 2, at 9; Alvaro Figueira & Bruno Vaz, *Survey on Synthetic Data Generation, Evaluation Methods and GANs*, MATHEMATICS, Aug. 2, 2022, at 1, 2.

32. Lu et al., *supra* note 2, at 9; Figueira & Vaz, *supra* note 31, at 2.

"Upsampling" methods aim to enhance a dataset's representation of minority subgroups,[33] thus bearing more resemblance to extrapolation. These methods often require more extensive (and sometimes more contestable) assumptions about ground truth. Returning to the Amazon example, suppose that Amazon sought to create synthetic resume data to counter the bias in its hiring algorithm by better representing women in the data it used to train the algorithm. Creating synthetic data to counter that bias is an exercise in gap-filling that requires making assumptions about the extent to which women were unfairly omitted from previous hiring decisions, the extent to which women applicants who should merit an interview will share characteristics with women or men who were previously hired, and so forth. Simulation methods (discussed below) can also be used to create synthetic data intended to fill known gaps in collected data, though they rely even more heavily on ground-truth assumptions. This shows, again, that whether augmented data can truly reduce the need for additional collected data depends on the reliability of the ground-truth assumptions.

In sum, the risks associated with using augmented synthetic data depend both on the collected data and on potentially contestable background knowledge and assumptions about the ground truth that the data represent. Because of this dual reliance, augmented synthetic data can be either more or less faithful to ground truth than the collected data it was based on, depending on the scope and representativeness of the collected dataset and on the validity of the ground-truth assumptions.

### 2. Simulated data

The third category encompasses synthetic data created by simulation methods, which *rely entirely on background knowledge and assumptions about ground truth* to generate synthetic data. It corresponds to Gal and Lynskey's category of generation "without (direct) use of collected data."[34] Data simulation models generate synthetic data through simulating real world conditions, rather than by modifying existing data.[35] For example, simulations of protein folding based on biological principles,[36] or of a game with clear rules, such as Go,[37] rely entirely on ground-truth assumptions about either the

33. Ludovico Boratto, Gianni Fenu & Mirko Marras, *Interplay Between Upsampling and Regularization for Provider Fairness in Recommender Systems*, 31 USER MODELING & USER-ADAPTED INTERACTION 421, 424 (2021); Sharma et al., *supra* note 12, at 359; Jaipuria et al., *supra* note 12, at 3344–45.

34. Gal & Lynskey, *supra* note 1, at 1100.

35. de Melo et al., *supra* note 5, at 175–77.

36. Josh Abramson et al., *Accurate Structure Prediction of Biomolecular Interactions with AlphaFold3*, 630 NATURE 493, 493 (2024).

37. David Silver, Thomas Hubert, Julian Schrittwieser & Demis Hassabis, *AlphaZero: Shedding New Light on Chess, Shogi, and Go*, GOOGLE DEEPMIND (Dec. 6, 2018), https://deepmind.google/discover/blog/alphazero-shedding-new-light-on-chess-shogi-and-go [https://perma.cc/3LE3-MH6T].

rules of the game or the properties of tangible objects to generate synthetic data. As Gal and Lynskey point out, simulations can also be based on assumptions about the statistical properties of the ground-truth distribution of relevant attributes, such as walking speed.[38] Simulation is especially helpful when testing and optimizing prior to physical experimentation or, perhaps most relevant to this discussion, when collected data is unavailable or too costly. For example, while it took researchers over fifty years to map roughly 100,000 protein structures, DeepMind's AlphaFold model predicted the structure of over 180 million proteins in the last several years.[39] The fidelity with which simulated synthetic data represents ground truth depends critically on the accuracy of the ground-truth assumptions embedded in the simulation.

Returning to the Amazon example once again, suppose that the company is concerned that the resumes of the women it has previously hired are unrepresentative of the ground-truth universe of potentially successful female hires. As a result, it cannot rely on straightforward manipulation or upsampling of the resume data it has at hand. One possibility might be to create a set of simulated resumes based on some assumptions about what the resumes of successful female hires would look like. This might be a difficult task, however. One might start by analyzing the resumes of previously successful male (or female) applicants. But what is the basis for assuming that the same sorts of backgrounds and experiences will predict success for other potential female applicants? To understand this, one might need to do a rather deep dive into what qualities make an employee successful, what experiences and characteristics are correlated with those qualities, and whether those experiences and characteristics differ by gender. One could go even further and question whether the qualities that make an employee successful now would stay the same in a more gender-balanced workforce. These are difficult questions, underlining the importance of ground-truth assumptions for creating simulated synthetic data.

### *C. Benefits of a Ground Truth–Focused Framework*

A ground truth-focused taxonomy of synthetic data helps one analyze the utility of various types of synthetic datasets, as well as where and how we might expect them to go wrong. For example, while some of the costs and benefits of synthetic data that Gal and Lynskey describe apply broadly (for example, the reduced cost of data labeling, curation, and perhaps storage), others apply

38. Gal & Lynskey, *supra* note 1, at 1101 ("Also interesting for our analysis are cases in which collected data is indirectly used, in that the simulation model relies on prior exposure of the coder (human or AI) to such data (i.e., background knowledge). The synthetic data is then based on the coder's assumptions regarding the statistical properties of the relevant data attributes. For example, the coder may base the maximum speed humans are shown reaching in synthetic videos on his real-world observations of human locomotion." (footnote omitted)).

39. John Jumper et al., *Highly Accurate Protein Structure Prediction with AlphaFold*, 596 NATURE 583, 583 (2021).

only in some (potentially narrow) situations in which the properties of the collected data, the validity of ground-truth assumptions, and the intended use of the synthetic data align. Indeed, many of the most exciting potential benefits of synthetic data depend heavily on how the data is created and what assumptions were made in creating it.

Transformed synthetic data aims to balance minimal loss of the characteristics of collected data with another goal, such as privacy. Transformation does not aim to improve how well the collected data captures ground truth; if anything, the transformation is likely to degrade that correspondence. Applications of transformed data can go awry if the collected data was inadequate to begin with or if the transformation does not properly balance fidelity to ground truth with other values.

Augmented synthetic data aims to enhance a collected dataset's correspondence to ground truth, whether the goal is to decrease bias or simply to obtain a larger dataset. Applications of augmented synthetic data can go awry if the augmentation relies on incorrect ground-truth assumptions, particularly if those mistakes exacerbate, rather than mitigate, weaknesses in the collected data.

Simulated synthetic data has various uses. It can be used to improve a collected dataset (as with augmented data) or to reduce reliance on collected data by replacing it, for example because collected data is costly or difficult to obtain. The reliability of simulated data depends particularly heavily on ground-truth assumptions. This reliance suggests that many successful applications of simulated data—whether to create entirely artificial datasets or to augment collected data—will involve situations where the relevant ground-truth assumptions are based on well-known principles such as the rules of a game or natural phenomena. Thus, it is unsurprising that many examples of simulated data are in arenas such as image generation.[40] It is also possible to analyze a sufficiently large and representative collected dataset to extract an approximation of the ground-truth principles that underlie the collected data, which could then be used to simulate synthetic data. In that approach, the extent to which simulated data represents ground truth will depend on how well the collected dataset represents ground truth.

The use of machine learning algorithms based on large datasets poses well-known problems of explainability and interpretability, particularly when used for decision-making. As scholars (including some of us) have argued in relation to AI-based decision-making tools, the focus should be on accountability, which can sometimes be achieved by requiring disclosures about factors such as data provenance—which in the case of synthetic data would mean the process by which it was produced.[41] Gal and Lynskey consider

40. Lu et al., *supra* note 2, at 2–6.

41. *See, e.g.*, Katherine J. Strandburg, *Adjudicating with Inscrutable Decision Rules*, *in* MACHINES WE TRUST: PERSPECTIVES ON DEPENDABLE AI 61, 61–62 (Marcello Pelillo & Teresa Scantamburlo eds., 2021); Ignacio N. Cofone & Katherine J. Strandburg, *Strategic Games and Algorithmic Secrecy*, how explainability and interpretability requirements should be applied to synthetic data.[42] Here also, recognizing the importance of ground-truth assumptions adds to the analysis. For synthetic data, disclosures of data provenance should include disclosure of the ground-truth assumptions that went into the production method. Moreover, just as with collected data, high-level disclosures regarding data provenance, choice of features, and the like may be insufficient in circumstances where a human decisionmaker or decision-subject needs—or has a right—to know why and how a specific decision was made.[43] In those cases, it may be necessary to use an explainable decision algorithm—or not to use any algorithm—whether or not synthetic data is used.[44]

Overall, one cannot fully analyze the potential costs and benefits of synthetic data from a policy perspective without considering how ground-truth assumptions affect the analysis. We illustrate, in the next Sections, the importance of following through on this emphasis by reconsidering the legal implications of synthetic data explored by Gal and Lynskey for privacy, data quality, and competition. We are persuaded by their main arguments and note that they have carefully acknowledged limitations to their analysis. We suggest, however, that additional insights can be gained by bringing our emphasis on ground-truth assumptions to bear on these questions.

## II. SYNTHETIC DATA AND PRIVACY

### A. *SYNTHETIC DATA AND PRIVACY LAW PRINCIPLES*

Discussions of synthetic data and privacy ordinarily emphasize the way that synthetic data can produce de-identified (loosely, "anonymous") data that is difficult to reidentify with individuals. The underlying idea is that synthetic data disconnects the distribution, behavior, and qualities of collected data from real individuals while preserving the use value of aggregated data.[45] Thus, as Gal and Lynskey argue, synthetic data can potentially promote key privacy law principles, such as "data security, data minimization, and data quality."[46] The role of ground-truth assumptions, along with the ways that privacy risk profiles differ among types of synthetic data, adds nuance to the common assessment of synthetic data's implications

64 MCGILL L.J. 623, 625 (2019); Ignacio Cofone & Katherine J. Strandburg, *Unjustifiable Algorithmic Opacity*, 15 N.Y.U. J. INTELL. PROP. & ENT. L. (forthcoming 2026).

42. Gal & Lynskey, *supra* note 1, at 1149.

43. *Id.* at 1150 ("[P]rocedural legitimacy might not be fit for all contexts in which synthetic data can be used. For instance, we might insist that where explainability is essential, synthetic data should only be used as training data but not as test data.").

44. *Id.* at 1150–51.

45. Fida K. Dankar & Mahmoud Ibrahim, *Fake It Till You Make It: Guidelines for Effective Synthetic Data Generation*, APPLIED SCIS., Feb. 28, 2021, at 1, 13; Liu et al., *supra* note 4, at 1379.

46. *See* Gal & Lynskey, *supra* note 1, at 1122.

for privacy, which extend beyond the question of reidentification of transformed data.

Potential leakage of identified or reidentified personal data is not the only significant privacy risk that arises when large datasets are used to draw inferences about individuals and groups. Privacy harms can also arise from intrusive inferences, which can be derived not only from transformed data, but also from augmented and simulated synthetic data. As the next Sections explain, the privacy benefits of synthetic data thus depend not only on the synthetic data's technical ability to prevent reidentification of individually identifiable collected data, but also on how and whether the synthetic data mitigates (or, depending on ground-truth assumptions, perhaps even exacerbates) a whole range of relevant privacy risks.

Moreover, as Gal and Lynskey discuss, transformed synthetic data is both a blessing and a curse as a matter of current privacy law: while transformation decreases the risk of reidentification, it may also take the data beyond the reach of most privacy laws.[47] In most legal regimes, only identifiable data are covered by privacy law, while anonymous (or sufficiently de-identified) data are not considered personal data.[48] Thus, while transformed synthetic data produces a certain level of technical protection against reidentification, its use may counterintuitively deprive individuals of legal protection. Augmented and simulated synthetic data are even less likely to be covered by most privacy laws because they are not directly derived from particular individuals' collected data. Whether the loss of legal protection is worth it depends on the category of synthetic data involved and the types of privacy harms at issue in a particular context.

### B. *Individual Data Harms: Reidentification and Leakage*

As noted, the most-discussed goal of using synthetic data to enhance privacy is de-identification (or "anonymization") of data within collected datasets. Enthusiasm for this prospect has risen due to the development of sophisticated de-identification approaches such as differential privacy. Not surprisingly, the most-noted *risk* of using synthetic data for privacy preservation is the possibility that a faulty transformation will allow for reidentification—a risk that Gal and Lynskey consider.[49] Reidentification occurs when someone (often an adversary) links synthetic data to real information about real individuals, undermining the privacy benefits of using

47. *See* Gal & Lynskey, *supra* note 1, at 1126–33. *But see* Ganev, *supra* note 3, at 1.

48. *See generally* Gilad L. Rosner, *De-Identification as Public Policy*, 3 J. DATA PROT. & PRIV. 250 (2020) (examining the lack of legally based requirements and incentives in Canada for de-identifying personal data under the Personal Information Protection and Electronic Documents Act while contrasting with the practices of different countries and suggesting a reduction in identification risks over handling the unfeasible difficulties of achieving anonymity). *See, e.g.*, Commission Regulation 2016/679, art. 2, 4(1), 2016 O.J. (L 119) 32–33 (EU).

49. *See* Gal & Lynskey, *supra* note 1, at 1125, 1138.

synthetic data.[50] This can happen if the synthetic data generation process retains too much information from the original collected data, allowing people to discern patterns and match the data to real individuals. Some data can be reidentified easily, such as location data.[51] In one study, just four location points throughout a year were enough to reidentify ninety-five percent of 1.5 million Belgian cellphone users.[52] Reidentified data are risky for individuals, as consequential material harms can accrue from them.[53] This combination of harms can take place, for example, when de-identified social security numbers are reidentified, exposing people to identity theft.[54]

The risk of leaking data about an identifiable individual depends on the category of synthetic data. For transformed synthetic data, the risk of reidentification can be high or low depending on the transformation method. And a transformation's robustness depends, among other things, on the ground-truth assumptions made, for example, about the relationships between various features in the data and the prevalence of particular data profiles in the population.[55]

This risk of reidentification is exacerbated if undue reliance is placed on an ineffective transformation method, either by privacy law as discussed above[56] or as a technical matter. The extent to which there is a risk of reidentification from various transformation methods has received considerable attention in the technical and legal literatures.[57] Because

50. *See* Theresa Stadler, Bristena Oprisanu & Carmela Troncoso, *Synthetic Data – Anonymisation Groundhog Day*, PROC. 31ST USENIX SEC. SYMP. 1451, 1451–52 (2022); Ziqi Zhang, Chao Yan & Bradley A. Malin, *Membership Inference Attacks Against Synthetic Health Data*, J. BIOMEDICAL INFORMATICS, 2022, at 1, 6–8 (reviewing highly susceptible vulnerabilities in synthetic health data to membership inference attacks).

51. Arvind Narayanan & Vitaly Shmatikov, *Myths and Fallacies of "Personally Identifiable Information*,*"* COMMC'NS ACM, June 2010, at 25–26; Melissa Gymrek, Amy L. McGuire, David Golan, Eran Halperin & Yaniv Erlich, *Identifying Personal Genomes by Surname Inference*, 339 SCI. 321, 324 (2012).

52. Larry Hardesty, *How Hard Is It to 'De-Anonymize' Cellphone Data?*, MASS. INST. OF TECH. (Mar. 27, 2013), https://news.mit.edu/2013/how-hard-it-de-anonymize-cellphone-data [https://perma.cc/DD85-QTLM].

53. Paul Ohm, *Broken Promises of Privacy: Responding to the Surprising Failure of Anonymization*, 57 UCLA L. REV. 1701, 1746–48 (2010).

54. Mark A. Geistfeld, *Protecting Confidential Information Entrusted to Others in Business Transactions: Data Breaches, Identity Theft, and Tort Liability*, 66 DEPAUL L. REV. 385, 385, 404 (2017).

55. *See* Da Zhong et al., *Disparate Vulnerability in Link Inference Attacks Against Graph Neural Networks*, 4 PROC. ON PRIV. ENHANCING TECHS. 149, 156–57 (2023) (finding that membership inference attacks are disproportionately successful against subgroups which are less densely represented in neural network); *see, e.g.*, Gymrek et al., *supra* note 51.

56. *See, e.g.*, Mason Marks & Claudia E. Haupt, *AI Chatbots, Health Privacy, and Challenges to HIPAA Compliance*, 330 JAMA 309, 309–10 (2023); I. Glenn Cohen & Michelle M. Mello, *Big Data, Big Tech, and Protecting Patient Privacy*, 322 JAMA 1141, 1141–42 (2019).

57. Cohen & Mello, *supra* note 56, at 1142; Woodrow Hartzog & Ira Rubinstein, *The Anonymization Debate Should Be About Risk, Not Perfection*, 60 COMMC'NS ACM 22, 22 (2017); Ira S. Rubinstein & Woodrow Hartzog, *Anonymization and Risk*, 91 WASH. L. REV. 703, 710–11 (2016);

synthetic data is designed to mimic the statistical properties of real data, synthetic data based on collected data can be too close to the original data—a problem similar the problem of overfitting in machine learning, where the AI corresponds to the training data too closely to generalize beyond that data.[58] This problem is likely to arise when the synthetic data includes rare combinations of attributes that were present in the original data but are not common in the population.[59] Thus, minority groups and outliers may be identifiable even after the collected data has been transformed. The most robust transformation techniques employ modern privacy-enhancing methods such as differential privacy or k-anonymity, which add noise or generalize data to make it difficult to reidentify individuals, while preserving certain statistical properties of the data and reducing the risk of overfitting.[60] Applying privacy-preserving techniques to synthetic data tends to have an outsized negative impact on accuracy for sub-groups underrepresented in the data, however, creating tradeoffs that may exacerbate unfairness.[61]

Augmented and simulated synthetic data mostly avoid the problem of reidentification since the synthetic data represents hypothetical individuals. Nonetheless, some techniques for data augmentation based on the collected data can, in principle, leak data about real individuals.[62] For example, augmented data made with upsampling techniques could create hypothetical profiles for a minority group that end up being too close to the small number of collected data profiles used to create them—and thus are identifiable to an individual or small group of individuals.

The extent of these leakage and reidentification risks depends on the relationship between the method, the collected data and the ground truth. Often there are trade-offs between fidelity to the properties of the collected data—which is of obvious importance for utility—and de-identification. Thus, methods that try to closely reproduce many properties of the collected data

---

Lisa M. Austin & Andrea Slane, *Digitally Rethinking* Hunter v. Southam, 60 OSGOODE HALL L.J. 421, 436 (2022).

58. Boris van Breugel, Hao Sun, Zhaozhi Qian & Mihaela van der Schaar, *Membership Inference Attacks Against Synthetic Data Through Overfitting Detection*, 206 PROC. 26TH INT'L CONF. ON A.I. & STAT. 3493, 3500–02 (2023) (preprint) (demonstrating a membership inference attack that "aims to infer membership by targeting local overfittinxvbg of the generative model").

59. Zhong et al., *supra* note 55, at 156–57; Stadler et al., *supra* note 50, at 1458.

60. *See* DWORK & ROTH, *supra* note 25, at 20–21. *See generally* Sweeney, *supra* note 26 (describing the k-anonymity model).

61. Georgi Ganev, Bristena Oprisanu & Emiliano De Crisofaro, Robin Hood *and* Matthew *Effects: Differential Privacy Has Disparate Impact on Synthetic Data*, 162 PROC. 39TH INT'L CONF. ON MACH. LEARNING 6944, 6949–50 (2022); Victoria Cheng, Vinith M. Suriyakumar, Natalie Dullerud, Shalmali Joshi & Marzyeh Ghassemi, *Can You Fake It Until You Make It?: Impacts of Differentially Private Synthetic Data on Downstream Classification Fairness*, PROC. 2021 ACM CONF. ON FAIRNESS, ACCOUNTABILITY, & TRANSPARENCY 149, 150 (2021).

62. Zhong et al., *supra* note 55, at 149–50; Stadler et al., *supra* note 50, at 1451–52.

have a higher risk of reidentification than methods, such as pure randomization, that allow for more deviation.[63]

In the next Section, we focus on a less discussed, but equally important, set of privacy risks posed by intrusive data-derived inferences. Unlike the risk of reidentification, these risks are not mitigated by synthetic data.

### *C. Harms from Group-Based Inferences*

Synthetic data has the potential to mitigate certain privacy harms related to data breaches and transfers, but the focus on individual reidentification harms is too narrow. As critics of current privacy law have pointed out in connection with anonymized data, privacy harms can arise not only from leakage of personally identifiable information, but also from privacy-intrusive, group-based inferences. Harms from group-based inferences can arise when large datasets, whether collected or synthetic, are used to derive inferences about preferences, behavior, population mobility, and so on.[64] All synthetic data can thus create inferential privacy risks because any of the three categories of synthetic data can be used to uncover group trends.[65] Even simulated data, which is generated from theoretical models or assumed distributions, can be used to derive intrusive group-based inferences. These privacy harms from group-based inferences exist in three forms.

First, privacy risks arise when an individual's personal data is input into a data-derived model to make privacy-intrusive inferences about them, regardless of whether that individual's data was included in the data used to train the model. Synthetic data does not ameliorate the privacy intrusion associated with using an individual's real data to make predictions or inferences about that individual. Consider the Amazon example to illustrate this point. Amazon could have trained its resume-selecting algorithm with simulated data, as opposed to data collected from Amazon employees. As soon as Amazon used the model, however, it would need to apply its algorithm to real personal data from job candidates to evaluate whether they would make desirable employees. The fact that the algorithm was trained on synthetic data

63. Nikhil Kandpal et al., *User Inference Attacks on Large Language Models*, 2024 PROC. CONF. ON EMPIRICAL METHODS NAT. LANGUAGE PROCESSING 18238, 18245–46 (demonstrating a user inference attack on LLMs based on users' prompt inputs during fine-tuning).

64. Alicia Solow-Niederman, *Information Privacy and the Inference Economy*, 117 NW. U. L. REV. 357, 423 (2022); Jordan M. Blanke, *The CCPA, "Inferences Drawn," and Federal Preemption*, 29 RICH. J.L. & TECH. 53, 58–60 (2022); Dara Hallinan & Frederik Zuiderveen Borgesius, *Opinions Can Be Incorrect (In Our Opinion)! On Data Protection Law's Accuracy Principle*, 10 INT'L DATA PRIV. L. 1, 9–10 (2020); Sandra Wachter & Brent Mittelstadt, *A Right to Reasonable Inferences: Re-thinking Data Protection Law in the Age of Big Data and AI*, 2019 COLUM. BUS. L. REV. 494, 497–98; Linnet Taylor, Bart van der Sloot & Luciano Floridi, *What Do We Know About Group Privacy?*, *in* GROUP PRIVACY: NEW CHALLENGES OF DATA TECHNOLOGIES 225, 231–32 (Linnet Taylor, Luciano Floridi & Bart van der Sloot eds., 2017).

65. Gal & Lynskey, *supra* note 1, at 1141–43.

does not mitigate the potential harm associated with collecting and using a candidate's real data to draw inferences about them.

Second, people can be harmed by synthetic data in the same ways that they are harmed by aggregated, de-identified information when it facilitates correct inferences about the groups they belong to and, by extension, about group members.[66] For example, if a company creates synthetic (transformed or augmented) data based on collected data about its users' sexual orientations, that synthetic data will preserve some statistical properties of the collected data. The company could then use the synthetic data to derive inferences about the preferences and behavior of queer individuals or to identify individuals as queer. The probabilistic information preserved by the synthetic data means that the data holder can infer something about queer individuals even if they were not included in the collected data. These privacy-intrusive inferences can expose people to harassment, discrimination, and human rights abuses, especially against members of vulnerable groups. The more inferences that the company can make, the more accurately it can target group members with personalized ads or algorithmic decision-making, which can produce additional harms such as manipulation or discrimination.[67] Gal and Lynskey discuss potential harms from overly precise inferences more generally in their discussion of data quality, on which we comment below.

Third, people can be harmed if biased synthetic data is used to make inaccurate inferences about them. While inaccurate inferences may not seem like privacy harms strictly speaking, they are data harms that privacy law ordinarily helps prevent and remedy.[68] Moreover, from the point of view of an individual, the intrusiveness of an inference may be exacerbated when the inference is incorrect, especially if it reflects bias or stereotypes. If the process for generating synthetic data is biased, the synthetic data can perpetuate biases present in the original collected data.[69] In particular, augmented data can amplify biases present in collected data. For example, if the original dataset contains racial or gender biases, data augmentation may extend these biases into the synthetic data because it is designed to mimic the statistical properties of the original dataset. When biased synthetic data is used to train a decision-making model, these biases will be imported into that model. Also, if outlier data is scarce, GANs can even exacerbate biases, for instance, by generating images of engineering professors that are more disproportionately masculine and fair-skinned than those in the original data.[70] Inaccuracy due to biased synthetic training data can be even more pernicious for marginalized groups than noise due to incomplete collected training data. This is because noise is a signal of unreliability, while a reduction in noise can obscure inaccuracies caused by bias and bolster the apparent reliability of a privacy-intrusive inference (e.g., privacy harms arising from intrusive inferences about potential pregnancies or LGBTQ2+ status).

---

66. Brent Mittelstadt, *From Individual to Group Privacy in Big Data Analytics*, 30 PHIL. & TECH. 475, 477–81, 485 (2017).

67. *See, e.g.*, Sandra Wachter, *Affinity Profiling and Discrimination by Association in Online Behavioral Advertising*, 35 BERKELEY TECH. L.J. 367, 369–75 (2020); Pauline T. Kim, *Race-Aware Algorithms: Fairness, Nondiscrimination and Affirmative Action*, 110 CALIF. L. REV. 1539, 1547–50 (2022).

68. Gal & Lynskey, *supra* note 1, at 1111–19; *see also* Ignacio Cofone, *Privacy Standing*, 2022 U. ILL. L. REV. 1367, 1396–97 (2022) (describing courts' considerations when identifying privacy harms).

69. *Cf.* Gal & Lynskey, *supra* note 1, at 1145–48 ("[Synthetic data] might reduce data quality when an analysis bases the generated synthetic data on incorrect assumptions.").

The level of this third risk depends on the type of synthetic data. Ironically, the risk from inaccurate privacy-intrusive inferences could be higher for augmented and simulated synthetic data, which are safer at the individual level of reidentification risk, because they are more susceptible to bad ground-truth assumptions. Augmented and simulated data can learn from biased historical data, such as hiring data with gender-biased practices in the Amazon case, encoding gender or racial discrimination into the assumptions used to create the synthetic data. Similarly, synthetic medical data may encode and preserve healthcare access disparities, and generated text may reflect harmful stereotypes present in web corpus data. Augmented and simulated data, in sum, can include synthetic data that is both lower quality and more biased than collected or transformed data. And inferences from that synthetic data may then be not only equally intrusive, but also more likely to be biased and inaccurate. Amplifying this risk, simulated data is, in many contexts (and perhaps increasingly), cheaper to produce than real collected data, making it easier and more tempting to create models based on questionable assumptions that exacerbate harmful biases.

Related to the harms from inferences based on synthetic data is the potential to generate synthetic data that closely resembles real individuals from entirely synthetic data, including simulated data.[71] Consider, for example, deepfakes. Deepfakes are simulated data but can be harmful to real individuals because of their relationship with ground truth: they are "fake" in one sense but individually harmful when constructed to resemble "fake" behavior from real people.[72]

These three types of group-based inference harms cannot be reduced by addressing privacy through standard legal mechanisms such as control rights and individual choices.[73] Privacy laws overlook possibilities to harm without

70. Niharika Jain, Alberto Olmo, Sailik Sengupta, Lydia Manikonda & Subbarao Kambhampati, *Imperfect ImaGANation: Implications of GANs Exacerbating Biases on Facial Data Augmentation and Snapchat Face Lenses*, A.I., Mar. 2022, at 1, 2–3.

71. Gal & Lynskey, *supra* note 1, at 1142.

72. *See, e.g.*, Shona Moreau & Chloe Rourke, *Fake Porn Causes Real Harm to Women*, POL'Y OPTIONS POLITIQUES (Feb. 8, 2024), https://policyoptions.irpp.org/magazines/february-2024/fake-porn-harm [https://perma.cc/9PUU-6K8E].

73. Przemysław Pałka, *Data Management Law for the 2020s: The Lost Origins and the New Needs*, 68 BUFF. L. REV. 559, 625–33 (2020) (arguing for inferred, non-synthetic data).

reidentification by ignoring these forms of group and individual harms.[74] How close synthetic data is to identifying an individual is thus sometimes the wrong question for privacy law. The right question is how likely the data is, regardless of its individual identifiability, to present risks for individuals or groups.

Privacy law, for that reason, should set standards to minimize privacy risks associated with both collected and synthetic data, rather than creating dichotomies focused on anonymization and reidentification risk.[75] Indeed, Gal and Lynskey note that formalist privacy laws that safeguard data about an individual struggle to protect them from harmful inferences facilitated by synthetic data or data about third parties.[76] However, they caution that the alternative might end up "casting the net" of privacy too widely, diminishing the utility of third party and publicly available data by subjecting it to data protection regimes.[77] The issue is not that synthetic data is in itself risky; it presents lower overall privacy risk than identified collected data. As Gal and Lynskey suggest, using synthetic datasets as opposed to collected data (often) complies with the principle of data minimization.[78] The deeper problem is that synthetic data demonstrates how privacy law currently operates under an unworkable binary of "personally identifiable" information versus "anonymous" information and deals inadequately with, or ignores, the privacy problems associated with group-based inferences.

No category of synthetic data, in sum, can prevent the often-overlooked harms stemming from intrusive group-based inferences (beyond individual reidentification attacks and data leakage from transformed data). And different types of synthetic data affect these risks differently depending on their relationship to ground truth in each particular context. Therefore, organizations generating or using synthetic data need to avoid treating it as data that inherently fulfills privacy principles and makes data safe. Instead, when generating or using it, they should implement these principles (as well as, when appropriate, conduct privacy impact assessments) as thoughtfully as they should do for collected data.

---

74. IGNACIO COFONE, THE PRIVACY FALLACY: HARM AND POWER IN THE INFORMATION ECONOMY 47–54 (2024).

75. *See* Hartzog & Rubinstein, *supra* note 57, at 22; Felix T. Wu, *Defining Privacy and Utility in Data Sets*, 84 U. COLO. L. REV. 1117, 1162 (2013) (making a similar argument for anonymous data).

76. Gal & Lynskey, *supra* note 1, at 1136, 1142–43 (presenting the concern).

77. *Id.* at 1143.

78. *Id.* at 1122–23 ("[B]y replacing collected data with artificially generated data . . . synthetic data offers an additional layer of security to personal data. Data minimization requires that only the minimum amount of personal data be processed for a specified purpose." (footnote omitted)).

## III. SYNTHETIC DATA AND DATA QUALITY

Gal and Lynskey's article discusses how synthetic data can increase the quality of a (collected) dataset, along with the potential effects of such increased quality on data power and social welfare and their implications for law and policy.[79] Focusing primarily on personal data, Gal and Lynskey highlight concerns that "a high-quality dataset could become a double-edged sword, as more accurate decisions might not always increase social welfare."[80] They explain that models trained on higher quality data could allow for more accurate personalization, which may be undesirable in some contexts.[81] For example, more accurately targeted advertising might lead to excessive price discrimination[82] or manipulative political messaging.[83] Personalized evaluation of tort law reasonableness or pretrial flight risk might be undesirable for rule-of-law values, and personalized risk assessments might undermine insurance markets.[84] As we note above, overly granular data-derived inferences can also create privacy harms. Thus, Gal and Lynskey argue that "the law should encourage those instances in which more accurate data-based decisions increase welfare, while prohibiting those in which it significantly reduces it."[85]

Gal and Lynskey define data quality in terms of "completeness"—which "ensures that certain data features are not unrepresented in a dataset"—and "accuracy"—which "ensures that they are not misrepresented in the dataset."[86] They clarify that by accuracy they follow data science usage to mean "the fraction of outputs of a model that are correct."[87] While "accuracy" is ordinarily measured in data science relative to a set of training or test data, in the context of synthetic data we take it to mean the fraction of model outputs that correspond to ground truth. Consequently, synthetic data would

79. *See id.* at 1143–53.

80. *Id.* at 1145–46.

81. *Id.* at 1146 ("[O]verly accurate information can enable new forms of differentiation and categorization, which might have negative welfare effects on individuals and groups through exploitation or manipulation.").

82. *Id.* (noting that "synthetic data's potential contribution to the creation of more accurate digital profiles" may cause an individual to "receive microtargeted offers for products that better fit their preferences, but possibly at higher, discriminatory prices that reflect their elasticity of demand").

83. *Id.* at 1147 (arguing that due to more accurate profiles created with synthetic data, "[i]n the political sphere, [an individual's] personalized digital feed could be designed to strengthen certain opinions and affect their political choices").

84. *Id.* (noting that with the personalization enabled by synthetic data, "digital profiles could inform decisions made by law enforcement or judicial bodies (e.g., based on a suspect's presumed flight risk), and even lead to the creation of personalized laws" (citing Omri Ben-Shahar & Ariel Porat, *Personalizing Negligence Law*, 91 N.Y.U. L. REV. 627 (2016))).

85. *Id.*

86. *Id.* at 1144 (emphasis omitted).

87. *Id.* at 1144 n.345 (citing Aileen Nielsen, Accuracy Bounding: A Regulatory Solution for the Algorithmic Society 6–9 (2022) (unpublished manuscript) (on file with author)).

contribute to "accuracy" if it allowed for training a model that is more representative of ground truth than available collected data. Synthetic data could thus strengthen data quality by augmenting collected data in ways which allow for the training of more accurate and more representative models. This possibility of enhanced quality refers most often to augmented data, and sometimes to simulated data, since transformation ordinarily reduces data quality in this sense.

### A. *Limitations on Synthetic Data's Capacity to Improve Data Quality*

Before raising questions about the relationship between synthetic data and excessive personalization, which is the key concern of Gal and Lynskey's discussion of data quality, it is worth emphasizing that synthetic data will not always improve data quality—and may even provide false assurance that quality problems with a collected dataset have been rectified. Gal and Lynskey mention that synthetic data can "reduce data quality when an analyst bases the generated synthetic data *on incorrect assumptions*."[88] Our emphasis on ground-truth assumptions brings this possibility to the fore.

Whether synthetic data can improve data quality depends on its category. Transformed data cannot improve the quality of a collected dataset, as these transformations forfeit some degree of data quality in exchange for improvements in other social values, such as privacy. Augmented and simulated data might improve dataset quality in comparison to (purely) collected data, but they are only as good as the ground-truth assumptions that go into them.

Picture two scenarios. In one, simulated data is based on well-established scientific or statistical principles (or perhaps the rules of a game), so it can be as accurate, or more so, than any data that could reasonably be collected due to cost or practicality. In the other, simulated data is based on questionable assumptions about ground truth, for example due to tentative background knowledge or generalizing from unrepresentative collected data. In that situation, the quality of the resulting dataset can be either better or worse than a collected dataset might be.

When adding synthetic data to augment nonrepresentative collected data, it is tempting to think that the synthetic data will always improve the dataset's completeness regarding cases that were underrepresented in the collected data (for example, minority subgroups of a population).[89] However, the accuracy of a machine learning model trained using an augmented dataset depends strongly on the validity of the ground-truth assumptions made about those missing cases and can vary systematically among

88. *Id.* at 1147 (emphasis added).

89. *See* Boratto et al., *supra* note 33, at 422–23; Sharma et al., *supra* note 12, at 359; Jaipuria et al., *supra* note 12, at 3344–45.

subgroups.[90] If the assumptions are wrong, then the apparent completeness of the augmented dataset can even be misleading as to the quality of the outputs, both overall and especially as to those subgroups.[91] Methods for interpolating between collected data can produce similar problems if they tend to smooth out kinks or outliers that represent ground-truth behaviors.[92] To give a simplified example, if I have three data points, it makes a big difference whether I assume they are drawn from a line, a parabola, or a triangle. While having more data points alleviates these concerns, similar issues recur when each case has many features, such that the possible functional relationships become exponentially more complicated.

Synthetic data may also fail to improve data quality if accuracy problems are due to missing features (characteristics) in the collected data. It might be that the collected dataset does not capture information that is important for creating an accurate model. Relevant features might be omitted for reasons of data availability or because of mistaken assumptions about what features are relevant.[93] In either case, unless the synthetic data generation method adds those missing features, its capacity to improve accuracy will be limited. Incorporating features that are not included in the collected data would be a nontrivial exercise in simulation that would require significant assumptions about ground truth. Similarly, if only a small number of cases from a sub-group are included in the collected data, one cannot assume that those cases are representative of the sub-group. Another way in which incorporating synthetic data might lead to lower data quality is if it is created using background assumptions that fail to update as situations on the ground evolve and change.[94] This can also happen if collected data goes out of date: the

90. The essential role of assumptions about outliers and missing cases is strongly supported by the existence of model collapse absent collected data about outliers, *see* Ilia Shumailov et al., *AI Models Collapse When Trained on Recursively Generated Data*, 631 NATURE 755, 755–56 (2024), and inference attacks against synthetic data based on detection of overfitting, *see* van Breugel et al., *supra* note 58, at 3499–500.

91. *See* Sierra Wyllie, Ilia Shumailov & Nicolas Papernot, *Fairness Feedback Loops: Training on Synthetic Data Amplifies Bias*, PROC. 2024 ACM CONF. ON FAIRNESS, ACCOUNTABILITY, & TRANSPARENCY 2113, 2119 (2024); *see also* Shumailov et al., *supra* note 90, at 755–56 ("Being trained on polluted data, [the models] then mis-perceive reality."); van Breugel et al., *supra* note 58, at 3499 (describing how the assumptions can "give[] a false sense of security" to underrepresented groups).

92. *See* Shumailov et al., *supra* note 90, at 757.

93. For instance, some researchers have proposed using a GAN to impute synthetic data for missing values and comparing those data to the original data, which requires making certain assumptions about similarities between the two. *See* Thomas Poudevigne-Durance, Owen Dafydd Jones & Yipeng Qin, *MaWGAN: A Generative Adversarial Network to Create Synthetic Data from Datasets with Missing Data*, ELECS., Mar. 8, 2022, at 1, 2.

94. Daniel Vela et al., *Temporal Quality Degradation in AI Models*, SCI. REPS., July 8, 2022, at 1, 1 ("[D]ata-producing environments, often change with time, and their statistical properties change alongside them. Known as 'concept drift', this evolution of data inevitably affects the quality of the models, to the point where the model may no longer correspond to its new reality." (footnotes omitted)).

need to update background assumptions could be overlooked, especially when those assumptions become baked into a data augmentation practice.

All of this is to say that, while synthetic data can often improve data quality in the sense Gal and Lynskey (as well as many others) intend, it is important to keep in mind its potential to exacerbate (or extend) the problems associated with collected data. When assumptions about ground truth are incorrect or insufficient, the resulting synthetic data can fail to improve a dataset or even make it worse. Assuming that augmented or simulated data always enhances a dataset can lead to over-confidence about the dataset's quality.[95] This is a particular concern where synthetic data is used as an inexpensive replacement for collecting more representative data, especially when the synthetic data is intended to represent personal information about social sub-groups.[96] When an under-represented group's ground truth is not sufficiently well understood or represented in the data, augmenting that same data can mislead and exacerbate existing biases.[97] Going back to the Amazon example of under-representation of females in the employee pool, it might or might not be correct to assume that women who were hired and performed successfully in the past are a representative sample of women who would perform successfully if hired today. The history of bias in hiring might have distorted the selection of women who were hired. As a result, there might also be insufficient background knowledge about what features would characterize women who would successfully fill the positions at issue.

### *B. Synthetic Data and Too Much of a Good Thing*

As Gal and Lynskey point out, previous scholars have argued that too much accuracy, especially in algorithms used for targeting and decision-making based on personal data, can reduce welfare.[98] These arguments generally refer to harms due to the ability to infer information about individuals in a privacy-intrusive manner or the ability to make granular inferences that permit manipulative or otherwise socially undesirable targeting.[99] The first kind of harm might include something like using purchase data to infer pregnancy or sexual orientation, as discussed above.[100] The second might include manipulative political or commercial advertising

95. Wyllie et al., *supra* note 91, at 2115–16.

96. *Id.* at 2114; Jain et al., *supra* note 70, at 5.

97. Jain et al., *supra* note 70, at 5.

98. Gal & Lynskey, *supra* note 1, at 1146–47.

99. *Id.*

100. Aziz Z. Huq & Rebecca Wexler, *Digital Privacy for Reproductive Choice in the Post-*Roe *Era*, 98 N.Y.U. L. REV. 555, 585–87 (2023); Isobel Cockerell, *Researchers Say Their AI Can Detect Sexuality. Critics Say it's Dangerous*, .CODA (July 13, 2023), https://www.codastory.com/authoritarian-tech/ai-sexuality-recognition-lgbtq [https://perma.cc/85WK-JQA9].

or overly granular price discrimination.[101] There are at least two senses in which increasing dataset quality might lead to these sorts of harms. First, the inclusion of more features in a dataset can allow greater granularity in algorithmic outputs, surfacing previously inaccessible categorical distinctions (between pregnant and non-pregnant consumers, for example). Second, having more or more reliable data can increase certainty about the algorithm's categorizations (even if they are not more granular), making algorithm users more likely to take (adverse) action based on the output.

Augmented datasets do not ordinarily add features to the collected dataset in ways that would permit increased granularity in the outputs because they generally use the features that are already in the collected data. The most likely way for synthetic data to increase accuracy by increasing granularity is by using simulated data that models features for which there is reliable background knowledge, but for which it is impossible or difficult to collect real world data. This scenario seems unlikely for the sorts of personal data algorithms that motivate the concern with over-granularity because features for which there is no collected data are often features for which the ground-truth relationships are poorly understood. Moreover, it may be hard to take action based on features that are generally missing from collected data. One can imagine exceptions. For example, a judge considering pretrial detention can ask the defendant for information that is not contained in a recidivism risk algorithm.

Most likely, then, any concerns that synthetic data would produce algorithms that are "too accurate" will arise when synthetic data increases certainty about an algorithm's categorizations without making them more granular. In other words, the concern is not with an increased number of output categories, but with more accurate placement of individuals into those categories leading to greater reliance on the categories. This is a possibility in principle, but its likelihood in practice depends on several things. First, it depends on whether the application is one in which algorithm users are likely to demand high certainty before acting. Such scruples seem intuitively less likely in the contexts of targeted advertising, whether commercial or political, and price discrimination, for example, than in the contexts of criminal sentencing or perhaps the setting of insurance rates. Second, it depends on how much the synthetic data increases confidence in the output. This is the flipside of the discussion above about whether synthetic data can make matters worse and depends on both the application and the validity of the ground-truth assumptions that are used in creating the synthetic data.

Finally, and most relevant to this discussion, Gal and Lynskey consider whether some laws should be adjusted to reflect how the use of synthetic data

101. Almog Simchon, Matthew Edwards & Stephan Lewandowsky, *The Persuasive Effects of Political Microtargeting in the Age of Generative Artificial Intelligence*, PNAS NEXUS, Feb. 2024, at 1, 3; Wachter, *supra* note 67, at 376–77.

changes the social values balance.[102] They first point out that many laws, such as those prohibiting deceptive practices or bias, should be unaffected, since they simply regulate harmful outcomes regardless of the underlying data.[103] They argue that the use of synthetic data might make a difference, however, for "laws that focus on data quality as a requirement for decision-making" which they argue generally assume "that improved data quality will increase social welfare."[104] This point harkens back to the concern that synthetic data might improve data quality and thereby create excessively accurate algorithms. Their references to medical insurance, in which the law balances the benefits of broad insurance coverage with insurers' desire to tailor coverage granularly to risks,[105] can help make the point. This example, along with proposals by Paul Ohm and by Aileen Nielsen,[106] bolsters Gal and Lynskey's suggestion that the law should require limits on algorithm accuracy in some situations.[107] Because the individual data used in making decisions presumably will not be synthetic, the rise of synthetic data should not affect data quality requirements of this sort.

Although legal limits on privacy-intrusive inferences and harmful over-personalization are important, the jury is still out on whether the rise of synthetic data has much bearing on when and how to set such limits. This is because synthetic data seems unlikely to support increased granularity in most situations. If the use of synthetic data leads to a new and unanticipated level of certainty about some potentially harmful, but yet unregulated, output, new law might be needed. It might make most sense to ban or regulate the harmful uses directly, though, rather than focus on the use of synthetic data.

Overall, while it is possible for the use of synthetic data to increase data quality in a way that triggers concerns about overly accurate algorithms, whether it does so in a particular case depends on the reliability of the ground-truth assumptions made in creating the synthetic data. The resulting

102. Gal & Lynskey, *supra* note 1, at 1150–54.

103. *Id.* at 1151 ("Take, for example, consumer protection laws which prohibit data-based deceptive practices, or laws that prohibit certain types of data-based bias, whether directly or through fairness requirements. Such laws apply based on the outcome and thus capture both real and synthetic datasets." (footnotes omitted)).

104. *Id.* at 1151–52.

105. *Id.* at 1152 ("The law often recognizes the merits of broad insurance coverage and limits the information that can be relied upon by insurers to calculate premiums. In this sense, the law acts as a constraint on accuracy, to promote a better power balance between the relevant parties and to achieve broader social goals." (footnote omitted)).

106. *Id.* at 1153 ("Ohm suggests the creation of 'throttling metrics,' by which friction in the algorithm might protect important human values, and Nielsen proposes that the accuracy of automated decision-making systems may be bounded where the output is too accurate for the context and leads to social harms." (footnotes omitted)).

107. As an aside, we note that many of the cited examples of laws that insist on highly accurate data, such as the Federal Privacy Act and the Fair Credit Reporting Act, regulate the accuracy of data about *specific individuals* used to make decisions *about those individuals*. *See id.* at 1151; 5 U.S.C. § 552a(e)(5) (2018); 15 U.S.C. § 1681e(b).

increased or decreased accuracy will affect social welfare in much the same ways as more or less accurate and complete collected data. Thus, while we agree with Gal and Lynskey's conclusion that synthetic data may "strengthen[] the case for regulation which focuses on usage," we are unconvinced that such regulation should be adopted *instead of* regulation focused "on data provenance"[108]—at least if data provenance is taken to include how the synthetic data was produced—because the use of synthetic data can increase or decrease algorithm accuracy depending on the validity of the ground-truth assumptions used to create it. Synthetic data should not be uniformly disfavored or given a pass. But it is critically important to interrogate its provenance in the sense of its basis in ground-truth assumptions.

## IV. COMPETITION AND SYNTHETIC DATA

Finally, Gal and Lynskey argue in their article that synthetic data can reduce the "durable market power" enjoyed by incumbents whose position (perhaps as result of network effects) affords them uniquely easy access to collected data.[109] In data-driven markets, potential competitors may face barriers to entry because they lack similar data access.[110] This situation can lead not only to higher consumer prices, but also to reduced innovation and suboptimal lock-in.[111] Gal and Lynskey argue that synthetic data can bolster competition "[b]y introducing an alternative to some types of collected data or by lowering the amounts of collected data needed" to compete, thus "augment[ing] collected datasets which are otherwise too small to be useful."[112] As a result, they argue that synthetic data will reduce incumbent firms' first-mover advantage and the incentives for mergers and market concentration.[113]

These potential competitive benefits derive from the possibility that competitors can use synthetic data to either augment or replace collected data; as a result, new competitors might get away with having less collected data. This argument applies to augmented and simulated synthetic data. Our emphasis on ground-truth assumptions is helpful in determining where and when such benefits are likely.

---

108. Gal & Lynskey, *supra* note 1, at 1154–55.

109. *Id.* at 1110–20.

110. *Id.* at 1111 ("[C]ompetition in data and data-based markets is shaped by the height of access barriers to data. When such barriers are high, potential entrants might not be able to challenge incumbents who enjoy data-based advantages because they cannot provide users with the utility that stems from better datasets." (footnote omitted)).

111. *Id.* at 1111–12 (noting that, due to data-based "durable market power," "productive and dynamic efficiency could be harmed because firms with potential cost or quality advantages might not be able to enter the market, and the incentives of incumbents to develop consumer-welfare-enhancing innovations could be suppressed.").

112. *Id.* at 1112.

113. *Id.*

In some fields, ground-truth assumptions, perhaps based on well-established or widely shared background knowledge, may be equally accessible to all competitors, regardless of access to collected data. To return to the example of protein folding, scientists can determine a protein's three-dimensional structure by drawing on foundational biochemical principles. While AlphaFold models greatly accelerated this process, they were trained on publicly available protein structures and the laws of nature.[114] In those contexts, simulated data can help overcome data-related barriers to entry. However, markets in which (1) data is expensive or impractical for new entrants to collect and (2) background assumptions are sufficiently well and widely understood for synthetic data to be competitive with collected data might be limited to contexts in which there is sufficient understanding of ground truth to permit simulation based on background knowledge, such as image production, protein folding, and the like.

Outside of such contexts, data augmentation is the more plausible approach for a smaller competitor to use, for example to substitute for collecting personal information from large numbers of users of social media or other applications. How likely is it, in such cases, that a competitor with a small dataset of collected user information can use synthetic data to compete effectively with an incumbent with a very large collected dataset? In our view, augmented datasets based on small amounts of collected data are unlikely to significantly "[r]educe the [n]eed for [c]ollected [d]ata"[115] of this sort. Even when they do, it is unclear that potential competitors would benefit much from such reductions. To begin with, any synthetic data technique that can be used by a potential competitor is also available to the incumbent to augment its own collected dataset. The incumbent also has several advantages in creating (more or more precise) synthetic data. Because the incumbent's dataset is, by assumption, very large, it is likely to be more representative of ground truth, making both interpolation- and extrapolation-type approaches more successful. For example, the incumbent's large dataset is more likely to contain examples of minority and outlier data that can be used as a basis for upsampling. Indeed, collected data about outliers may become increasingly important because excessive use of synthetic data in AI training can risk causing "model collapse," in which the tails of the original content distribution disappear.[116] The incumbent also is more likely to be able to use its collected data to derive principles that can be used to create better simulated data.

---

114. Abramson et al., *supra* note 36, at 495.

115. Gal & Lynskey, *supra* note 1, at 1098.

116. Shumailov et al., *supra* note 90, at 755; Mohamed El Amine Seddik, Suei-Wen Chen, Soufiane Hayou, Pierre Youssef & Merouane Debbah, *How Bad is Training on Synthetic Data? A Statistical Analysis of Language Model Collapse*, CONF. ON LANGUAGE MODELING, Apr. 7, 2024, at 1, 11 (finding "that model collapse always happens when the model is training solely on synthetic data" but can be avoided by mixing real and synthetic data).

There will be some situations in which a smaller collected dataset combined with synthetic data based on that dataset provides similar market performance to a larger collected dataset (and, potentially, synthetic data based on that larger dataset). For instance, collecting additional data might sometimes provide diminishing returns once synthetic data is feasible and a smaller competitor might be able to collect enough data to reach that point. We fear that situations in which augmented data enables a smaller competitor to catch up with the incumbent's advantages are likely to be rare, however. It seems more likely that the incumbent's ability to combine its large, collected dataset with its advantages in synthetic data creation will ordinarily allow it to maintain its position. That will be particularly likely when the market is also characterized by network effects that are not data-dependent, as is the case, for example, for social media applications.

Another competition issue involves mergers and acquisitions aimed at aggregating data. The availability of synthetic data promises to decrease motivations for mergers and acquisitions of similar companies simply to aggregate "more" data of a similar sort. All data is not equal, however. Many of these mergers seem aimed at combining *heterogeneous* datasets, involving either different types of data (e.g., Google/DoubleClick)[117] or data from different swaths of society (e.g., Facebook/Instagram).[118] The UnitedHealth/Change HealthCare merger, similarly, appears to have been motivated by a desire for data of a sort that was not previously accessible to UnitedHealth.[119]

Assuming there are no widely known ground-truth principles that could be used to simulate the "other" set of data, it seems unlikely that synthetic data based on a company's current collected data would provide the heterogeneity that often seems to have motivated mergers aimed at aggregating collected datasets. Analysis of the role of ground-truth assumptions thus suggests that, for many potential mergers, synthetic data is unlikely to replace the collected data held by the other company, and hence unlikely to decrease incentives for mergers.

The extent to which the possibility of creating synthetic data will render "obsolete" regulations such as "mandatory data sharing, portability, interoperability, and standardization"[120] raises similar questions. How often and in what markets synthetic data can eliminate barriers to entry (or eliminate the need for interoperability) is an empirical question. The

117. Gal & Lynskey, *supra* note 1, at 1116.

118. *Id.*

119. Gal & Lynskey suggest UnitedHealth sought the merger to get "access to deidentified health data on millions of patients, enabling it to cherry-pick the most profitable geographic areas" in which to enhance insurance coverage. *Id.* at 1117–18.

120. Gal & Lynskey, *supra* note 1, at 1092, 1121 ("[N]ew regulations are being suggested, and cases are being brought, based on assumptions of data-based market power that are no longer true for some markets . . . Recognizing the effects of synthetic data may lead to a more nuanced, hands-off regulatory approach in this legal realm.").

importance of ground-truth assumptions for creating reliable synthetic data shows that we do not yet know whether "the introduction of synthetic data will lead to a less interventionist approach to data-based advantages that affect competition."[121] The emergent use of synthetic data will almost certainly shift the playing field for competition, but to what extent and in which situations remains an open question.

## CONCLUSION

Foregrounding the role of background knowledge and assumptions about ground truth in synthetic data is essential to fully understand the legal and policy implications of synthetic data. Our Comment builds on Gal and Lynskey's incisive analysis of the ways in which synthetic data will revolutionize information privacy, data quality, and market competition. Although we agree with many of Gal and Lynskey's arguments, some of the advantages and disadvantages that they identify may apply only in narrower contexts based on the type of synthetic data at issue, its relationship to collected data, the validity of its creators' ground-truth assumptions, and its intended use.

Gal and Lynskey examine when the law should incentivize or mandate the use of synthetic data. We wholeheartedly agree with their argument that synthetic data is not an all-purpose fix for problems of bias and discrimination, but instead a potentially reasonable approach that should be encouraged or even mandated in certain contexts.[122] Our contribution to this discussion is to emphasize that synthetic data should be used when its underlying ground-truth assumptions can be sufficiently justified. Thus, an inquiry into the validity of those assumptions should be part of any consideration of whether to incentivize the use of synthetic data.

Overall, we argue that policy discussions about synthetic data should emphasize the relationship between different categories of synthetic data and ground-truth assumptions. As a result, we propose classifying synthetic data as (1) transformed data, which modifies collected data based on assumptions about which of the collected dataset's statistical properties should be preserved for a given end use; (2) augmented data, which relies on ground-truth assumptions to augment a collected dataset in order to improve its fidelity to ground truth for a given purpose; and (3) simulated data, which relies heavily on background knowledge and assumptions about ground truth. We believe that reframing the taxonomy of synthetic data in terms of reliance on ground-truth assumptions adds important insights to Gal and Lynskey's seminal analysis of the legal and policy implications of synthetic data.

---

121. *Id.* at 1121.

122. *Id.* at 1148–49 (noting that "synthetic data should not be treated as a quick or even the most efficient fix for all illegal data-based decisions" but that "where synthetic data can be relatively easily and cost effectively used to reduce illegal harms, it should be taken into account by courts").